\documentclass[aps,twocolumn,showpacs,prr,amsmath,amssymb,floatfix,superscriptaddress,longbibliography]{revtex4-2}
\usepackage{amsmath,amssymb,bm}
\usepackage[dvips]{graphicx}
\usepackage{yfonts}
\usepackage{enumerate}
\usepackage{color}
\usepackage{comment}
\usepackage{here}
\usepackage[
colorlinks=true,
pdfstartview=FitV,
linkcolor=blue,
citecolor=magenta,
urlcolor=blue,
bookmarks=true,
bookmarksnumbered=true,
pdftitle={},
pdfauthor={}
]{hyperref}

\begin{document}

\author{Yuki Minami} \email{yminami@keio.jp}
\affiliation{Department of Physics and Zhejiang Institute of Modern Physics, Zhejiang University, Hangzhou, Zhejiang 310027, China}
\author{Gentaro Watanabe} \email{gentaro@zju.edu.cn}
\affiliation{Department of Physics and Zhejiang Institute of Modern Physics, Zhejiang University, Hangzhou, Zhejiang 310027, China}
\affiliation{Zhejiang Province Key Laboratory of Quantum Technology and Device, Zhejiang University, Hangzhou, Zhejiang 310027, China}

\date{\today}

\title{Effects of pairing gap and band gap on superfluid density in the inner crust of neutron stars}
\begin{abstract}
Calculations of the superfluid density in the inner crust of neutron stars by different approaches are in strong disagreement, which causes a debate on the accountability of pulsar glitches based on superfluidity. Taking a simple unified model, we study the dependence on approximation of the superfluid density in a periodic potential. In comparison with the Hartree-Fock-Bogoliubov (HF-Bogoliubov) theory which treats the effects of the band gap and the pairing gap on equal footing, we examine the HF-BCS-type approximation in which the former is incorporated in priority, and another approximation in which the latter is incorporated in priority. We find that, when the pairing gap and the band gap are comparable as in the inner crust of neutron stars, they need to be treated on equal footing, and the HF-BCS approximation can considerably underestimate the superfluid density even if the pairing gap is much smaller than the Fermi energy. Our result suggests that the validity of the HF-BCS approximation for evaluating the superfluid density in neutron star crusts is questionable.
\end{abstract}

\maketitle

\section{Introduction}
Neutron stars are very accurate natural clocks spinning at a regular period. However, it is also known that their spinning rate sometimes suddenly increases and then relaxes to that before the spin-up. This phenomenon is called “glitch”, which is a long-standing problem of neutron star physics. A promising model of glitches is based on neutron superfluidity in the crust of neutron stars~\cite{sauls1989superfluidity, pethick1995matter,  haskell2015models, sedrakian2019superfluidity, andersson2021superfluid}. In the inner crust, nuclei form a crystalline lattice, and neutrons permeating the lattice forms a superfluid which possesses quantized vortices~\cite{shapiro2008black, haensel2007neutron}. It is considered that, while the vortices are usually pinned to the nuclei consisting of a normal fluid, the sudden collective unpinning of the vortices releases the angular momentum of the superfluid to the normal fluid, which is observed as a glitch~\cite{anderson1975pulsar, pines1980pinned, pines1985superfluidity, epstein1988vortex, PhysRevLett.110.241102, PhysRevLett.117.232701}. The neutron superfluid density is a fundamental quantity for understanding glitches since it is directly related to the strength of the glitches.

Although the superfluid can flow without friction, it can be still entrained by the lattice of nuclei mainly due to the band gap caused by the periodic lattice. To estimate the entrainment effect, Chamel~\cite{chamel2012neutron} has performed a three-dimensional band calculation of neutrons in the inner crust based on the Hartree-Fock (HF) approximation. He has obtained the superfluid density that is about one tenth of the believed value. Such a low value of the superfluid density is not large enough to explain the observed strength of the glitches, and the standard glitch model based on neutron superfluidity in the crust runs into serious difficulties. Consequently, the entrainment effect on the superfluid density attracts great interests~\cite{PhysRevC.94.065801, watanabe2017superfluid,  chamel2018, kashiwaba2019self, sauls2020superfluidity, sekizawa2021time}.

However, the pairing effect is neglected in the HF approximation and its importance has been pointed out in Ref.~\cite{watanabe2017superfluid}. The equal-footing treatments of the pairing gap and the band gap leads to recovery of the superfluid density that is around $70$ percent of the believed value. On the other hand, Chamel~\cite{chamel2018} reported an updated calculation showing that the superfluid density is still low even if the pairing effect is included by the BCS approximation. Therefore, the problem of the superfluid density in neutron star crusts is still controversial. The disagreement between these studies suggests that the superfluid density depends on the approximation method whether or not the pairing gap and the band gap are included on equal footing.

There is a recent development in the study of the entrainment effect on the slab phase of neutron star matter~\cite{kashiwaba2019self, sekizawa2021time}. The slab phase has one-dimensional periodic layers of nuclei like lasagna pasta~\cite{ravenhall1983structure, hashimoto1984shape} and provides a simple setup for calculations. A self-consistent band calculation without pairing in Ref.~\cite{kashiwaba2019self} yields the reduction of the effective mass of neutrons in the slab phase by a factor of $0.65$--$0.75$ from their bare mass. In addition, a time-dependent self-consistent band calculation in Ref.~\cite{sekizawa2021time} has also reported the reduction of the collective mass of neutrons. In the latter work, the reduction of the mass is interpreted to be caused by a counter flow of the dripped neutrons along the opposite direction to the motion of the slab.
These studies claim that the entrainment effect enhances the mobility of the dripped neutrons and the conventional scenario of the glitch phenomena based on neutron superfluidity is still tenable for the slab phase. However, the slab phase is a thin layer of the inner crust and the interpretation of the glitch is still unclear for the entire inner crust.

Another recent direction is on the impurity disorder of the lattice~\cite{jones1999amorphous, sauls2020superfluidity}. An effect of structural disorder on neutron superfluidity in the crust has been studied using a model of an amorphous metallic alloy~\cite{sauls2020superfluidity}. The resulting superfluid density is sufficient to explain glitch phenomena in contrast to that of the regular lattice.

In the present paper, we will shed light on the dependence of the superfluid density on the approximation methods and the importance of the equal-footing treatment of the pairing and the band mixing. The previous results have been obtained by large-scale self-consistent band calculations, and thus it is not clear what causes the differences among these results. Therefore, in the present work, we take a simple minimal model of a fermionic superfluid under a periodic potential with minimal three bands including the pairing $\Delta$ and the band mixing $V$.
With this minimal model, we compare the following three methods: (i) We directly solve the Bogoliubov--de Gennes (BdG) equations with $V$ and $\Delta$ on equal footing, which is so-called the HF-Bogoliubov theory. (ii) We first incorporate only the band mixing by $V$, and then include the particle-hole mixing by $\Delta$. This method can be regarded as the HF-BCS approximation~\cite{broglia2013fifty}. (iii) We first incorporate only the particle-hole mixing by $\Delta$, and then include the band mixing by $V$. Methods (ii) and (iii) are approximations of Method (i). We shall demonstrate that the resulting superfluid density shows a significant difference among the three methods when $\Delta \sim V$ even if $\Delta$ is much smaller than the Fermi energy $E_F$ as in the case of superfluid neutrons in the inner crust. Especially, the superfluid density by Method (ii) can be considerably smaller than that by Method (i) in such a situation. We also discuss the implication of our result to the superfluid density in the inner crust of neutron stars.

This paper is organized as follows. We present our minimal model in Sec.~\ref{sec:model}, and explain the details of the three methods to calculate the superfluid density in Sec.~\ref{sec:method}. In Sec.~\ref{sec:results}, we present the results showing that there is a considerable difference in the superfluid density among the three methods when $\Delta \sim V$ even if $\Delta \ll E_F$. Implications of our result to the superfluid density of the inner crust of neutron stars are discussed in Sec.~\ref{sec:discussion}. Finally, concluding remarks are given in Sec.~\ref{sec:concluding remarks}.

\section{Model}
\label{sec:model}

The BdG equations in one dimension with the superfluid momentum $Q$ are written as~\cite{watanabe2017superfluid, watanabe2008equation}
\begin{align}
\begin{pmatrix}
H_Q'(x) & \Delta(x) \\
\Delta(x) & -H_{-Q}'(x)
\end{pmatrix}
\begin{pmatrix}
u_i(x) \\
v_i(x)
\end{pmatrix}
=
\epsilon_i
\begin{pmatrix}
u_i(x) \\
v_i(x)
\end{pmatrix},
\label{BdG}
\end{align}
where $i$ is the shorthand index for the Bloch momentum and the band indices, $u_i$ and $v_i$ are the quasiparticle amplitudes of state $i$, $\epsilon_i$ is the corresponding quasiparticle energy, $\Delta(x)$ is the pairing gap, and $H'_Q(x)$ is given by
\begin{align}
H'_Q(x)&= \frac{1}{2m}\biggl(-i \frac{\partial}{\partial x} +Q\biggr)^2+V_{\rm ext}(x)- \mu(Q).
\end{align}
Here, $m$ is the bare neutron mass and $\mu(Q)$ is the chemical potential.
We choose the external potential $V_{\rm ext}$ with the following sinusoidal form:
\begin{align}
V_{\rm ext}(x)&=V\biggl(e^{iKx}+e^{-iKx}\biggr),
\end{align}
where the magnitude $V$ of the periodic potential is a real constant and $K$ is the reciprocal lattice vector. We self-consistently determine the $Q$-dependence of $\mu$ such that $n(Q)=n(Q=0)$, where $n(Q)$ is the particle density at $\mu(Q)$. We consider the periodic solutions such that the quasiparticle amplitudes $u_k(x)$ and $v_k(x)$ for quasimomentum $k$ are given by the Bloch states:
\begin{align}
u_k (x)=\sum_{j=0,\pm1}  \tilde{u}_{k+j K}e^{i(k+jK)x}, \label{bloch u}\\
v_k (x)=\sum_{j=0,\pm1}  \tilde{v}_{k+j K}e^{i(k+jK)x}, \label{bloch v}
\end{align}
where we restrict within three bands $j=0, \pm1$ for simplicity.

Substituting Eqs.~(\ref{bloch u}) and (\ref{bloch v}) into Eq.~(\ref{BdG}), we get
\begin{align}
\begin{pmatrix}
\tilde{H}'(k, Q) & \tilde{\Delta} \\
\tilde{\Delta} & -\tilde{H}'(k, -Q)
\end{pmatrix}
\begin{pmatrix}
\tilde{\bm{u}}_k \\
\tilde{\bm{v}}_k
\end{pmatrix}
=
\epsilon_k
\begin{pmatrix}
\tilde{\bm{u}}_k \\
\tilde{\bm{v}}_k
\end{pmatrix},
\label{BdG}
\end{align}
where $\tilde{\bm{u}}_k=(\tilde{u}_{k+K}, \tilde{u}_{k},\tilde{u}_{k-K})$, $\tilde{\bm{v}}_k=(\tilde{v}_{k+K}, \tilde{v}_{k},\tilde{v}_{k-K})$, and $\tilde{\Delta}={\rm diag}(\Delta, \Delta, \Delta)$. Here, we have neglected the momentum dependence of  $\Delta(k)$. $\tilde{H}'(k,Q)$ in the diagonal blocks is given by
\begin{align}
\tilde{H}'(k,Q)=&
\begin{pmatrix}
\xi(k+K,Q) & V & 0 \\
V & \xi(k,Q) & V \\
0 & V & \xi(k-K,Q)
\end{pmatrix}
\label{HF}
\end{align}
with
\begin{align}
 \xi(k,Q) &=\frac{1}{2m}(k+Q)^2-\mu(Q). \label{xi}
\end{align}
Equations (\ref{BdG}), (\ref{HF}), and (\ref{xi}) constitutes our minimal model with three Bloch bands. The explicit matrix form of our model is given in Appendix~\ref{sec:matrix}. In this simple model, both the strength $V$ of the external periodic potential characterizing the band gap and the strength $\Delta$ of the particle-hole mixing characterizing the pairing gap are introduced as pre-determined parameters.

\section{Methods}
\label{sec:method}

The superfluid density $n^{s}$ is defined as the second-order derivative of the energy density $e$ with respect to $Q$:
\begin{align}
n^{s}=\frac{\partial^2 e}{ \partial Q^2}\bigg\vert_{Q=0}.\label{def of ns}
\end{align}
To obtain the superfluid density, we have to calculate the energy density.
The energy density of the system described by the BdG equations is given by
\begin{align}
e = \int^{K/2}_{-K/2} \frac{d k}{2\pi} \sum_i \biggl[2(\mu-\epsilon_i)|v_i|^2+ \Delta^* u_i v_i^* \biggr].\label{energy density}
\end{align}
Here, we take a summation of $i$ over the positive eigenvalue modes. In Method (i), we straightforwardly calculate the energy density (\ref{energy density}) with the solutions of Eq.~(\ref{BdG}). The superfluid density by this method is denoted by $n^s_{\rm BdG}$. Method (i) is so-called the HF-Bogoliubov theory which incorporates $V$ and $\Delta$ on equal footing, and provides the reference result to examine Methods (ii) and (iii), which are further approximations of Method (i). Below, we explain the energy density and the superfluid density by Methods (ii) and (iii).

\subsection{Method (ii)}
First, we consider only the band mixing by $V$ and solve the eigenvalue problems for $\tilde{H}'(k,Q)$ and $-\tilde{H}'(k,-Q)$ in Eq.~(\ref{BdG}), respectively:
\begin{align}
\tilde{H}'(k,Q) \bm{\psi}^j(k,Q) &= \epsilon^j_{p,V}\bm{\psi}^j(k,Q),  \label{eq:HFevp}\\
\tilde{H}'(k,-Q) \bm{\psi}^j(k,-Q) &= \epsilon^j_{h,V}\bm{\psi}^j(k,-Q),\label{eq:HFevh}
\end{align}
where $j$ labels the band, and the subscripts $p$ and $h$ in the eigenvalue represent the ``particle'' and ``hole'', respectively. This procedure corresponds to solving the HF equations for a periodic potential and $\bm{\psi}^j(k,\pm Q)$ corresponds to the HF basis.

\begin{figure*}[t]
  \begin{center}
    \begin{tabular}{c}
      \begin{minipage}{0.5\hsize}
        \begin{center}
          \includegraphics[width=\hsize]{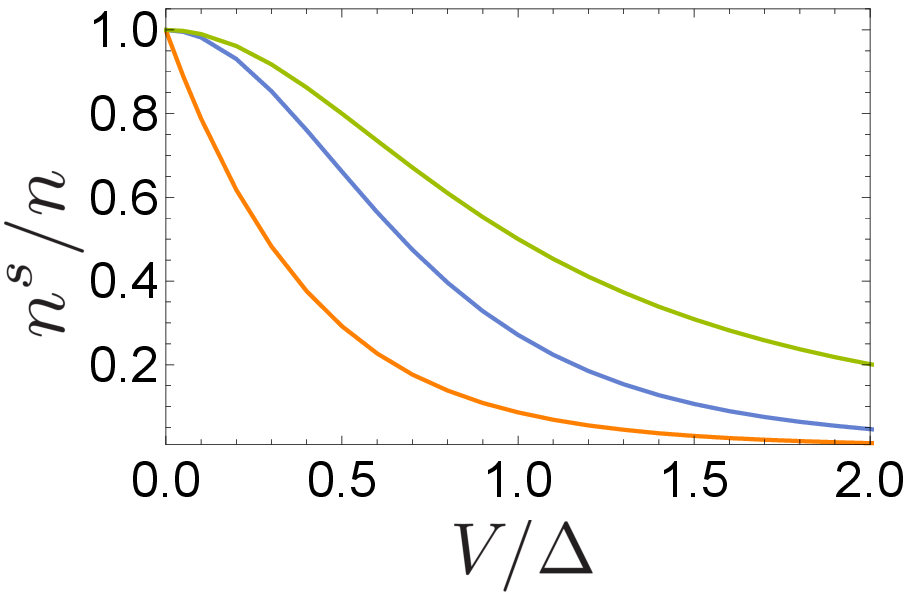}
        \end{center}
      \end{minipage}
            \begin{minipage}{0.5\hsize}
        \begin{center}
          \includegraphics[width=\hsize]{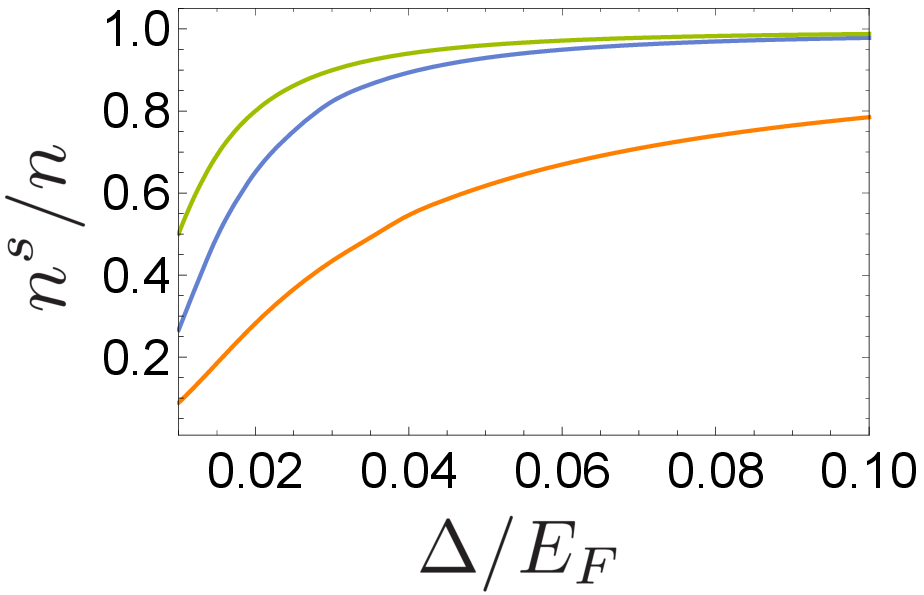}
        \end{center}
      \end{minipage}
    \end{tabular}
    \caption{Superfluid density $n^s$ by the three methods. The left panel shows the superfluid density as a function of $V/\Delta$ at $\Delta/E_F=10^{-2}$ and $K/k_F=2$. The right panel shows the superfluid density as a function of $\Delta/E_F$ at $V/E_F=10^{-2}$ and $K/k_F=2$. Since the numerical accuracy is low for very small $\Delta/E_F$, we plot down to $\Delta/E_F = 0.1$ in the right panel. The blue, orange, and green lines show $n^s_{\rm BdG}/n$, $n^s_{V \Delta}/n$ and $n^s_{\Delta V}/n$, respectively.
    }
    \label{fig:ns1}
  \end{center}
\end{figure*}

Next, we consider the particle-hole mixing by the pairing gap $\Delta$, and solve the following BdG equations:
\begin{align}
\begin{pmatrix}
\epsilon^j_{p,V}, &  \Delta \\
\Delta & -\epsilon^j_{h,V}
\end{pmatrix}
\begin{pmatrix}
u_j \\
v_j
\end{pmatrix}
= \epsilon_{V\Delta}^j
\begin{pmatrix}
u_j \\
v_j
\end{pmatrix}.
\label{BdG VD}
\end{align}
We calculate the energy density (\ref{energy density}) with the solutions of Eq.~(\ref{BdG VD}). The superfluid density obtained by this method is denoted by $n^s_{V\Delta}$. As we shall discuss in Sec.~\ref{sec:HFBCS}, Method (ii) can be identified as the HF-BCS approximation employed in~\cite{chamel2018}. This approximation incorporates the band gap in prior to the pairing gap.

\subsection{Method (iii)}
In Method (iii), we first incorporate only the particle-hole mixing due to $\Delta$ by solving the following BdG equations:
\begin{align}
\begin{pmatrix}
\xi_{k+jK, Q}, &  \Delta \\
\Delta & -\xi_{k+jK, -Q}
\end{pmatrix}
\begin{pmatrix}
\tilde{u}_{k+jK}^a \\
\tilde{v}_{k+jK}^a
\end{pmatrix}
= \varepsilon_{k+jK}^{a}
\begin{pmatrix}
\tilde{u}_{k+jK}^a \\
\tilde{v}_{k+jK}^a
\end{pmatrix}
 \label{Delta}
\end{align}
with $\xi_{k+jK, Q}=\xi(k+jK, Q)$, the band index $j=\{0, \pm 1\}$, and $a=+$ or $-$ which labels positive and negative eigenvalues, respectively. Next, we incorporate the band mixing due to $V$ by solving the following one-particle Schr\"odinger equation:
\begin{align}
H&
\begin{pmatrix}
\tilde{\psi}_{k+K}^l \\
\tilde{\psi}_{k}^l \\
\tilde{\psi}_{k-K}^l
\end{pmatrix}
=\epsilon_{\Delta V}^l
\begin{pmatrix}
\tilde{\psi}_{k+K}^l \\
\tilde{\psi}_{k}^l \\
\tilde{\psi}_{k-K}^l
\end{pmatrix}, \label{eq:Sh eq Delta to V}
\end{align}
with
\begin{align}
H&=\begin{pmatrix}
\mu-\varepsilon_{k+K}^- & V & 0 \\
V & \mu-\varepsilon_{k}^+ & V \\
0 & V & \mu-\varepsilon_{k-K}^-
\end{pmatrix}, \label{H}
\end{align}
where the index $l$ of the eigenvalue takes $0$, $1$, and $2$ for each $j$ band, and $l=0$ labels the lowest eigenvalue.

We have determined the diagonal elements of $H$ given by Eq.~(\ref{H}) by the following argument. The diagonal elements of $H$ are the energies of the modes of $k+jK$ ($j=\{0, \pm 1\}$) without the band mixing. From the expression of the energy density of the system (\ref{energy density}), $ (\mu - \epsilon_i)|v_i|^2+\Delta^* u_i v_i^*/2$ can be regarded as the energy of each mode $i$. In the region of $\Delta \ll \mu$, we can approximate that $|v_i|^2 \sim 1$ and $u_i v_i^* \sim 0 $ for the positive (negative) eigenvalue modes of $j=0$ ($j=\pm1$) band (see Appendix.~\ref{vsquare} for details). Therefore, we have set $\mu-\varepsilon_{k+K}^- $, $\mu-\varepsilon_{k}^+ $, and $\mu-\varepsilon_{k-K}^- $ for the diagonal elements of $H$ in Eq.~(\ref{H}).

From the energy eigenvalues $\epsilon_{\Delta V}^l$ obtained from Eq.~(\ref{eq:Sh eq Delta to V}), we calculate the energy density of the system as follows. We suppose that the $N$-particle wave function $\Psi$ is given by the Slater determinant of one-particle Bloch waves $\psi_k$ with the quasimomentum $k$ as:
\begin{align}
\Psi(x_1, x_2, ..., x_N)=\frac{1}{\sqrt{N !}}\det [\psi_{k_1}(x_1)\psi_{k_2}(x_2)...\psi_{k_N}(x_N)].
\end{align}
The energy of the lowest state is calculated as
\begin{align}
E_{\Delta V} &=\langle \Psi^0| H | \Psi^0 \rangle\nonumber \\
&=\sum_{n=1}^N  \int dx\,  \tilde{\psi}_{k_n}^{0}(x)^* H \tilde{\psi}^0_{k_n}(x). \label{isum}
\end{align}
Since the Hamiltonian~(\ref{H}) does not have a two-body interaction, the total energy of the system is written as a sum of the single-particle energy. We approximate the summation about $n$ in Eq.~(\ref{isum}) by an integration of the momentum $k$ over the Fermi sea $V_{\rm FS}$ with the Fermi momentum $k_F$ as follows:
\begin{align}
E_{\Delta V}
&=2 \sum_{k \in V_{\rm FS}}  \int dx\, \tilde{\psi}_{k}^{0}(x)^* H \tilde{\psi}^0_{k}(x)\nonumber \\
&= \frac{L}{\pi} \int^{k_F}_{-k_F} dk  \int dx\,  \tilde{\psi}_{k}^{0}(x)^* H \tilde{\psi}^0_{k}(x)\nonumber  \\
&= \frac{L}{\pi} \int^{k_F}_{-k_F} dk\,  \epsilon_{\Delta V}^0.
\end{align}
with $\epsilon_{\Delta V}^0(k) \equiv \int dx\, \tilde{\psi}_{k}^{0}(x)^* H \tilde{\psi}^0_{k}(x)$. Dividing by the system size $L$, we get the energy density
\begin{align}
e_{\Delta V}
&= \frac{1}{\pi} \int^{k_F}_{-k_F} dk  \epsilon_{\Delta V}^0.
\label{eq:eDeltaV}
\end{align}
The superfluid density $n^s_{\Delta V}$ by this method is given by
\begin{align}
n^s_{\Delta V}=\frac{\partial^2 e_{\Delta V}}{\partial Q^2}\biggl\vert_{Q=0}.
\end{align}
Opposite to Method (ii), Method (iii) incorporates the pairing gap in prior to the band gap.

\section{Results}
\label{sec:results}

The left panel of Fig.~\ref{fig:ns1} shows the superfluid density calculated by the three methods as a function of $V/\Delta$. Since the pairing gap of superfluid neutrons is typically the order of $1$\% of the neutron Fermi energy in the inner crust of neutron stars, here we set $\Delta/E_F=0.01$. In addition, since we examine the three methods for the situation in which the band gap effect is nonnegligible, we set $K/k_F=2$ where the Fermi surface is perfectly nested~\footnote{In the present minimal model with only three bands, effects of the band gap is negligible unless $K/k_F=2$.}.
Consequently, in our setup, the lowest band is almost filled and a part of the next lowest band within $\Delta$ from the Fermi energy is slightly filled.
For one-dimensional systems, the particle density $n$ is given by $n=2k_F/\pi$. The blue, orange, and green lines show the superfluid density by Methods (i), (ii), and (iii), respectively. The results show that the superfluid density by the three methods considerably differ in the region of $V \sim \Delta$ even though the pairing is negligibly small compared to the Fermi energy, $\Delta \ll E_F$. For example, at $V / \Delta=1$, we obtain
\begin{align}
n^s_{\rm BdG}/n &=0.27, \label{nsBdG} \\
n^s_{V\Delta}/n &=0.089, \label{nsHFBCS}\\
n^s_{\Delta V}/n &=0.50.
\end{align}
The superfluid density $n^s_{\rm BdG}$ by Method (i) shows an intermediate value between $n^s_{V \Delta }$ [Method (ii)] and $n^s_{\Delta V}$ [Method (iii)]. It is close to $n^s_{\Delta V}$ in the small $V/\Delta$ region and $n^s_{V \Delta}$ in the large $V/\Delta$ region.

Furthermore, in the region of small $V/\Delta$, the behavior of the superfluid density $n^s_{V \Delta}$ by Method (ii) is qualitatively different from the others. $n^s_{\rm BdG}$ and $n^s_{\Delta V}$ behave quadratically in $V/\Delta$ whereas $n^s_{V \Delta}$ behaves linearly. The reason of this difference can be understood as follows. Let us first consider a calculation only with the band mixing, which corresponds to the first half step of Method (ii). Our setup is the nested case, $K=2k_F$, and the Fermi surface (i.e., two points at $k=\pm k_F$) lies inside the band gap when $V$ is nonzero. As a consequence, the system without pairing is an insulator and the flow density is $n^s/n=0$ for any nonzero $V$. On the other hand, for $V=0$, there is no band gap and thus all the particles contribute to the flow, so that $n^s/n=1$ even without pairing. Therefore, $n^s$ shows step-wise behavior such that $n^s/n =1$ at $V=0$ and $n^s/n=0$ otherwise, and this behavior can be regarded as the reason of the qualitative difference of the superfluid density by Method (ii) from the others. In Method (ii), while this step-wise behavior is smoothened by the pairing effect incorporated in the second half step in Method (ii), the resulting superfluid density $n^s_{V \Delta}$ is still influenced by the step-wise behavior before including the pairing.

The right panel of Fig.~\ref{fig:ns1} shows the superfluid density as a function of $\Delta/E_F$ for $V/E_F=0.01$. The blue, orange, and green lines again show the superfluid density by Methods (i), (ii), and (iii), respectively. The superfluid density $n^s/n$ by both Methods (i) and (iii) rapidly increases and saturates to unity as $\Delta/E_F$ increases. Consequently, the difference between them is negligibly small for larger $\Delta/E_F \agt 0.1$. On the other hand, $n^s/n$ by Method (ii) slowly increases and does not saturate yet even at $\Delta/E_F=0.1$. In the limit of $\Delta/E_F \rightarrow 0$, the superfluid density by all the three methods should converge to zero while we cannot calculate for very small $\Delta/E_F$ below $\sim 0.01$ due to the limitation of the numerical accuracy.

\begin{figure*}[t]
  \begin{center}
    \begin{tabular}{c}
      \begin{minipage}{0.5\hsize}
        \begin{center}
          \includegraphics[width=\hsize]{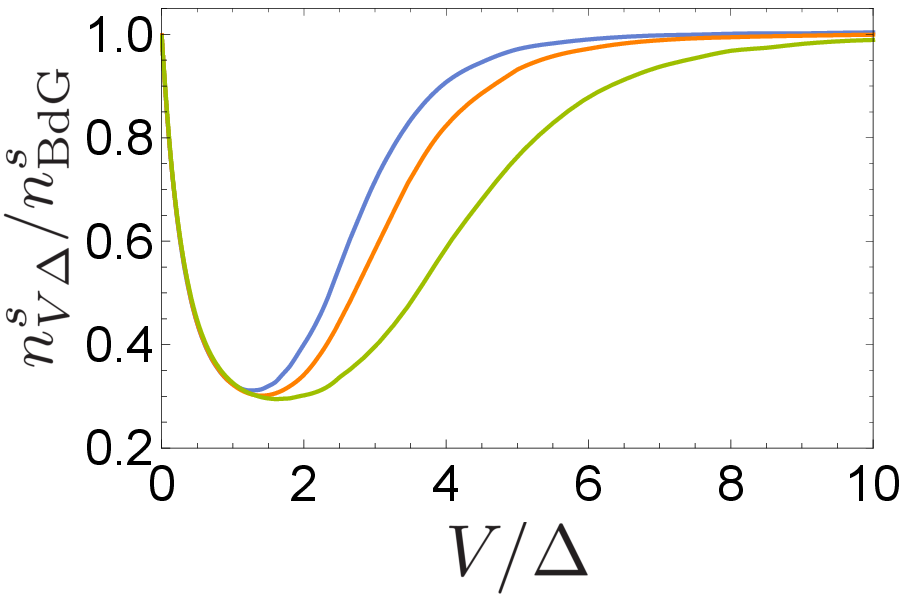}
        \end{center}
      \end{minipage}
            \begin{minipage}{0.5\hsize}
        \begin{center}
          \includegraphics[width=\hsize]{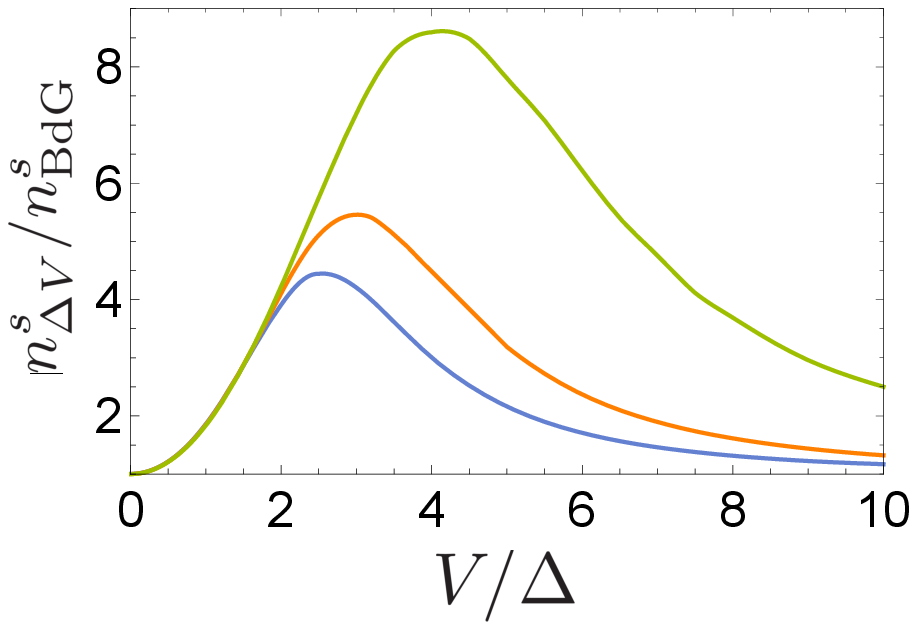}
        \end{center}
      \end{minipage}
    \end{tabular}
    \caption{Superfluid density $n^s_{V\Delta}$ by Method (ii) and  $n^s_{\Delta V}$ by Method (iii) in comparison with  $n^s_{\rm BdG}$ by Method (i). The left (right) panel shows the ratio $n^s_{V\Delta}/n^s_{\rm BdG}$ ($n^s_{\Delta V} / n^s_{\rm BdG}$) as a function of $V/\Delta$ at $K/k_F=2$ for several values of $\Delta/E_F$. The green, orange, and blue lines are for $\Delta/E_F=0.01$, $0.05$ and $0.1$, respectively.}
    \label{fig:ratio}
  \end{center}
\end{figure*}

In Fig.~\ref{fig:ratio}, we plot  the ratios $n^s_{V \Delta}/n^s_{\rm BdG}$ (left panel) and $n^s_{\Delta V}/n^s_{\rm BdG}$ (right panel) as functions of $V/\Delta$ for several values of $\Delta$. The green, orange, and blue lines show these ratios for $\Delta/E_F=0.01$, $0.05$, and $0.1$, respectively. For, e.g., $\Delta/E_F=0.01$ (green lines), $n^s_{V\Delta}/n^s_{\rm BdG}$ takes a minimum value of $0.29$ at $V /\Delta = 1.6$ (left panel) and $n^s_{\Delta V}/n_s^{\rm BdG}$ takes a maximum value of $8.6$ at $V/\Delta =4.0$ (right panel). This result clearly demonstrates that the resulting superfluid density can considerably differ by the approximation when the pairing gap and the band gap are comparable. Interestingly, this difference is more prominent when the pairing gap is negligibly smaller than $E_F$ (e.g., the green lines for $\Delta/E_F = 0.01$). As $\Delta/E_F$ increases, the minimum and maximum values of these ratios approach unity due to the saturation of $n^s/n$ as seen in the right panel of Fig.~\ref{fig:ns1}. In the region of $V /\Delta \alt 1$, the three lines converge to a single line and thus $\Delta/E_F$-dependence of these ratios becomes negligible.

Furthermore, the ratios $n^s_{V \Delta}/n^s_{\rm BdG}$ and $n^s_{\Delta V}/n^s_{\rm BdG}$ approach to unity in the limits of $V/\Delta \to 0$ and $V/\Delta \to \infty$ in Fig.~\ref{fig:ratio}. In the limit of $V/\Delta \to 0$, the reduction of the superfluid density by the band gap effect vanishes. Thus, all the resulting superfluid densities by the three methods agree with the particle density, so that $n^s_{V \Delta}/n^s_{\rm BdG} = n^s_{\Delta V}/n^s_{\rm BdG} = 1$. In the limit of $V/\Delta \to \infty$, on the other hand, the band mixing by $V$ is dominant and thus it is considered that the interplay between the effects of the band gap and the pairing gap is lost. As a consequence, once the effect of $V$ is included, the resulting superfluid density is unchanged irrespective of the order in which $V$ and $\Delta$ are included. In conclusion, the equal-footing treatment of $V$ and $\Delta$ by Method (i) is crucial for the reliable value of $n^s$ when $V$ and $\Delta$ are comparable. It is noted that they are indeed comparable in the inner crust of neutron stars as we will discuss in Sec.~\ref{sec:concluding remarks}.

\section{Discussion}
\label{sec:discussion}
\subsection{Correspondence to the HF-BCS approximation}
\label{sec:HFBCS}

We show that Method (ii) is essentially the same as the HF-BCS approximation employed in Refs.~\cite{carter2005effect, chamel2018}. We first diagonalize the diagonal blocks of $\tilde{H}'(k, Q)$ and $-\tilde{H}'(k, -Q)$ in Eq.~(\ref{BdG}). Here, we obtain two HF basis functions $\bm{\psi}(k,Q)$ and  $\bm{\psi}(k,-Q)$ defined in Eqs.~(\ref{eq:HFevp}) and (\ref{eq:HFevh}) for the particle Hamiltonian and the hole Hamiltonian, respectively. We also obtain the two orthogonal matrices $\tilde{\mathcal{O}}(k,Q)$ and $\tilde{\mathcal{O}}(k,-Q)$ which diagonalize each of them, and introduce the following matrix:
\begin{align}
\mathcal{O}=
\begin{pmatrix}
\tilde{\mathcal{O}}(k,Q) & O \\
O &\tilde{\mathcal{O}}(k,-Q)
\end{pmatrix},
\end{align}
where $O$ is a $3 \times 3$ zero matrix.
We rotate the BdG equations (\ref{BdG}) by $\mathcal{O}$, and obtain
\begin{align}
\begin{pmatrix}
\tilde{H}'_V(k, Q) & \tilde{\Delta}_V(k,Q) \\
 \tilde{\Delta}^*_V(k,Q)  & -\tilde{H}'_V(k, -Q)
\end{pmatrix}
\begin{pmatrix}
\tilde{\bm{U}} \\
\tilde{\bm{V}}
\end{pmatrix}
=
\epsilon
\begin{pmatrix}
\tilde{\bm{U}} \\
\tilde{\bm{V}}
\end{pmatrix}.
\end{align}
Here, $\tilde{H}'_V(k, Q)$ and $\tilde{H}'_V(k, -Q)$ are the diagonalized particle and hole Hamiltonians given by
\begin{align}
\biggl(\tilde{H}'_V(k, Q)\biggr)_{ij} &=\epsilon_{p,V}^i \delta_{ij}, \\
\biggl(\tilde{H}'_V(k, -Q)\biggr)_{ij} &=\epsilon_{h,V}^i \delta_{ij},
\end{align}
where $\epsilon_{p,V}^i$ and $\epsilon_{h,V}^i$ are the eigenvalues of Eqs.~(\ref{eq:HFevp}) and (\ref{eq:HFevh}), respectively.

The rotated pairing block is given by
$ \tilde{\Delta}_V(k,Q) =\tilde{\mathcal{O}}^{-1}(k,Q)\tilde{\Delta}\tilde{\mathcal{O}}(k,-Q)$.
We note that $\tilde{\mathcal{O}}^{-1}(k,Q)$ is not the inverse of $\tilde{\mathcal{O}}(k,-Q) $ for nonzero $Q$, and the rotated pairing block $\tilde{\Delta}_V(k,Q)$ has off-diagonal elements and depends on $Q$ and $k$.
In the HF-BCS approximation, we discard these off-diagonal elements:
\begin{align}
\biggl(\tilde{\Delta}_V(k,Q)\biggr)_{i j}
\longrightarrow \delta_{ij} \biggl(\tilde{\Delta}_V(k,Q)\biggr)_{i i},
\end{align}
which could be a reasonable approximation when the system is almost uniform.
Furthermore, in \cite{chamel2018, chamelPC}, the diagonal elements $(\tilde{\Delta}_V(K,Q))_{i i}$ for all $i$ are replaced by a single value $\Delta$, so that their dependence on $k$ and $Q$ are neglected. This treatment is essentially the same as Eq.~(\ref{BdG VD}) in Method (ii).

\subsection{Implications to  three-dimensional systems}
We discuss implications of our results to the superfluid density in the inner crust of neutron stars. Our model is in one dimension while the real system is in three dimensions. The reduction of the superfluid density by the band gap is the most prominent in the one-dimensional, perfectly nested case of $K=2k_F$ considered in the present work as we provide two reasons below. Therefore, our result of $n^s/n$ can be regarded as the most conservative value in which the suppression by the band gap is strongest: in the neutron star crusts, such suppression of the superfluid density should be weaker.

The first point is that the superfluid flow $\bm{Q}$ and the reciprocal lattice vector $\bm{K}$ can take various directions in three dimensions. The effect of the band gap suppressing $n^s$ should be weaker for a flow whose $\bm{Q}$ is not along $\bm{K}$. The actual periodic lattice potential in the inner crust of neutron stars has many Fourier components with various directions and magnitudes of $\bm{K}$. On the other hand, in one dimension, there is no orthogonal component of $\bm{K}$ to $\bm{Q}$, and thus the reduction of the superfluid density is expected to be the most prominent.

\begin{figure}[t!]
        \begin{center}
          \includegraphics[width=\columnwidth]{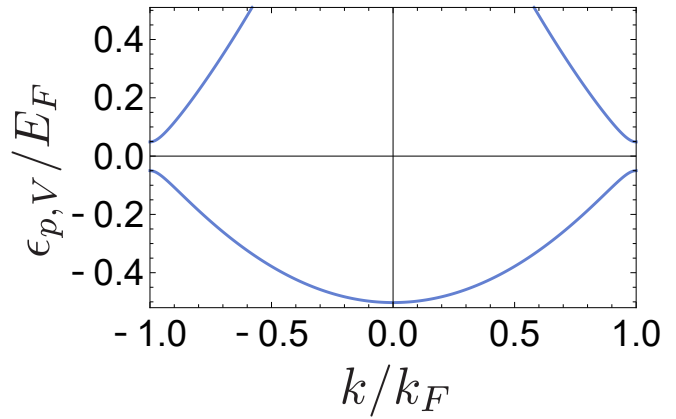}
        \end{center}
            \caption{Energy band structure of $\epsilon_{p, V}$ obtained from Eq.~(\ref{eq:HFevp}) for $K/k_F=2$ and $V/E_F=0.1$. The slope of $\epsilon_{p, V}$ is zero at ACPs on the Fermi surface at $k=\pm k_F$.
    }
    \label{fig:band}
\end{figure}

\begin{figure}[t!]
        \begin{center}
          \includegraphics[width=\columnwidth]{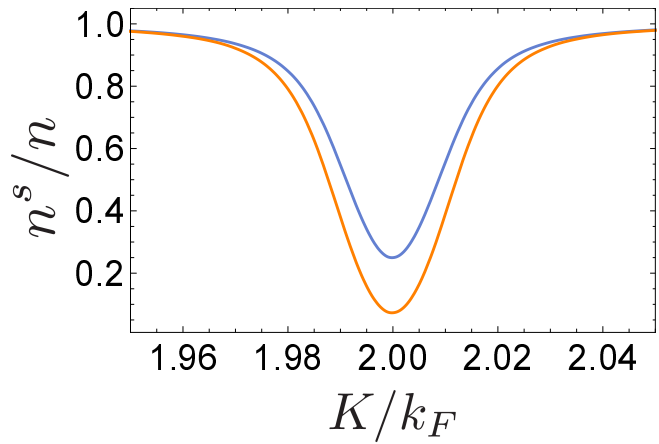}
        \end{center}
            \caption{Superfluid density $n^s$ as a function of $K/k_F$ at $\Delta/E_F=V/E_F=10^{-2}$. The blue and orange lines are for Methods (i) and (ii), respectively.
    }
    \label{fig:nsK}
\end{figure}

Next, the reduction of the superfluid density is caused by the increase of the effective mass of the dripped neutrons around avoided-crossing points (ACPs) in the band structure. Curvature of the energy dispersion is the inverse of the effective mass, and the slope of the dispersion is zero at the ACP as shown at $k/k_F = \pm 1$ in Fig.~\ref{fig:band}. To get large reduction of the superfluid density, almost all of the particles near the Fermi surface have to feel the effect of the ACPs. The Fermi surface in the one-dimensional momentum space is just two points at $k=\pm k_F$ and lie on the ACP when $K=2k_F$ as shown in Fig.~\ref{fig:band}. It is noted that the Fermi surface in three dimensions is a two-dimensional surface while the ACPs are points. Although three-dimensional systems have many reciprocal lattice vectors, they are just a finite number of points and cannot completely cover the two-dimensional Fermi surface. On the other hand, the Fermi surface in a one-dimensional system consists of two points, and the reciprocal lattice vectors completely overlap with these two points in the nested case. Therefore, the effect of the ACPs is the most prominent in our one-dimensional setting. Figure~\ref{fig:nsK} shows that the superfluid density of our model as a function of $K/k_F$. The superfluid density rapidly recovers as the Fermi surface moves away from the ACP. This result suggests that, in three-dimensional systems, the suppression of the superfluid density by the band structure is reduced compared to the one-dimensional systems due to the large part of the Fermi surface uncovered by the ACPs.

\section{Summary and Concluding remarks}
\label{sec:concluding remarks}

There is a serious disagreement among previous works~\cite{chamel2012neutron, watanabe2017superfluid, chamel2018} in the superfluid density in neutron star crusts, and this problem is still controversial. To get a better understanding of this issue, we have studied the dependence on approximation of the superfluid density in a periodic potential using a minimal model including the key competing effects of the band gap $V$ and the pairing gap $\Delta$. We have demonstrated three calculation methods: (i) We directly solve the BdG equations with $V$ and $\Delta$ on equal footing. (ii) We first incorporate only the band mixing by $V$, and then include the particle-hole mixing by $\Delta$.  (iii) We first incorporate only the particle-hole mixing by $\Delta$, and then include the band mixing by $V$. Methods (ii) and (iii) are not equivalent to Method (i), but they are different approximations of the latter such that Method (ii) incorporates $V$ in prior to $\Delta$ and Method (iii) incorporates $\Delta$ in prior to $V$. Method (i) is called the HF-Bogoliubov theory employed in \cite{watanabe2017superfluid} and Method (ii) corresponds to the HF-BCS approximation employed in \cite{chamel2018}. We have found that the resulting superfluid density is considerably different among the three methods when $V$ is comparable to $\Delta$ even if $\Delta$ is negligibly smaller than $E_F$. For example, in the perfectly nested case of $K/k_F=2$ in our one-dimensional minimal model, the superfluid density by Method (ii) is about one third of that by Method (i) for $\Delta= V = 0.01 E_F$ [Eqs.~(\ref{nsBdG}) and (\ref{nsHFBCS})].

Our results imply a possibility that the HF-BCS approximation considerably underestimates the superfluid density in the inner crust of neutron stars. In the inner crust, the pairing gap is estimated to be comparable to the magnitude of the lattice potential~\cite{watanabe2017superfluid}. Namely, the lattice potential has many Fourier components $V_K$ with wave vectors $K$ corresponding to the reciprocal lattice vectors, and the magnitude $|V_K|$ for the primitive reciprocal lattice vector is largest whose value is around $1$--$2$ MeV~\cite{pearson2015role, chamelPC, watanabe2017superfluid}. On the other hand, $^1S_0$ pairing gap in the density region of the inner crust is typically $\sim 1$ MeV~\cite{Gezerlis2014}. Therefore, we conclude that the equal-footing treatment of the pairing and the band mixing by the HF-Bogoliubov theory is essential to calculate the superfluid density of the inner crust in neutron stars. The HF-BCS approximation tends to underestimate the superfluid density, especially under the condition of the neutron star inner crust.

In the density region of the inner crust, where the superfluid density is drastically reduced in the calculation by Chamel \cite{chamel2012neutron}, typical values of the Fermi energy of dripped neutrons and the distance between the neighboring nuclei are $15$~MeV and $40$--$50$~fm, respectively. Therefore, the typical magnitude of the ratio between $K$ and $k_F$ is $K/k_F \sim 10^{-1}$. This value is very different from the value $K/k_F=2$ taken in our calculations. The real system allows an infinite number of bands and many ACPs exist around the Fermi surface, so that the band gap effect is nonnegligible even at $K/k_F \sim 10^{-1}$ as in Refs.~\cite{chamel2012neutron, watanabe2017superfluid}. On the other hand, since our minimal model has only three bands and the band gap is much smaller than the Fermi energy in the current problem, the effect of the band gap is negligible in our model unless $K/k_F=2$. To discuss the band gap effect as well as the pairing gap effect by this minimal model, we thus have set $K/k_F=2$.

In our results, the superfluid density is always smaller than the particle density whereas some previous studies on the slab phase reported that the effective mass of the neutrons is reduced from their bare mass~\cite{kashiwaba2019self, sekizawa2021time}. In our work, we consider permeating neutrons under the external periodic potential while these works have studied the system consisting of free neutrons, lattice neutrons (i.e., neutrons bound in nuclei), and protons with the Bloch boundary conditions. In the latter case, there is an ambiguity in the definition of the free neutrons~\cite{kashiwaba2019self}. The reduction of the effective mass suggests that some of the lattice neutrons turn to the conducting neutrons by the flow. Our setup does not have such an ambiguity, and the superfluid density is necessarily smaller than the particle density.

To understand the discrepancies in the superfluid density among previous calculations, we have employed a simple toy model. Once the difference among the approximations has been clarified within the minimal model in the present work, it is now interesting to study the approximation dependence using a model taking account of more realistic setup of the inner crust such as the lattice structure in three dimensions~\cite{Kobyakov2014}, defects and disorder of the lattice~\cite{DeBlasio1998, PhysRevC.105.055807}, realistic nuclear forces~\cite{Gandolfi2015} and so on. We leave these issues for future works.

\bigskip
\begin{acknowledgments}
We thank C.~J.~Pethick for fruitful discussions and N.~Chamel for providing the details of the calculation in~\cite{chamel2018}. Meetings in A3 Foresight Program supported by JSPS are also acknowledged for discussions. This work is supported by NSF of China (Grants No. 11975199 and No. 11674283), by the Zhejiang Provincial Natural Science Foundation Key Project (Grant No. LZ19A050001), and by the Zhejiang University 100 Plan.
\end{acknowledgments}

\bibliography{neutronstar_rev}

\appendix

\section{The explicit form of the BdG equation~(\ref{BdG})}
\label{sec:matrix}
The explicit matrix expression of Eq.~(\ref{BdG}) is as follows:
\begin{widetext}
\begin{align}
\begin{pmatrix}
\xi(k+K,Q) & V & 0 & \Delta & 0 & 0\\
V & \xi(k,Q) & V & 0 & \Delta & 0\\
0 & V & \xi(k-K,Q) & 0 & 0 &\Delta \\
\Delta & 0 & 0 & -\xi(k+K,-Q) & V & 0 \\
0 & \Delta & 0 & V & -\xi(k,-Q) & V  \\
0 & 0 &\Delta & 0 & V & -\xi(k-K,-Q)
\end{pmatrix}
\begin{pmatrix}
\tilde{u}_{k+K} \\ \tilde{u}_{k} \\ \tilde{u}_{k-K} \\ \tilde{v}_{k+K} \\ \tilde{v}_{k} \\ \tilde{v}_{k-K}
\end{pmatrix}
=\epsilon_k
\begin{pmatrix}
\tilde{u}_{k+K} \\ \tilde{u}_{k} \\ \tilde{u}_{k-K} \\ \tilde{v}_{k+K} \\ \tilde{v}_{k} \\  \tilde{v}_{k-K}
\end{pmatrix}\,.
\end{align}
\end{widetext}

\section{Behaviors of $|v|^2$ in the limit of $\Delta \to 0$}
\label{vsquare}
As a  simple example, we consider the following BdG equations:
\begin{align}
\begin{pmatrix}
\epsilon &  \Delta \\
\Delta & -\epsilon
\end{pmatrix}
\begin{pmatrix}
u \\ v
\end{pmatrix}
=
\lambda
\begin{pmatrix}
u \\ v
\end{pmatrix},
\end{align}
with $|\epsilon| \gg |\Delta|$.
The eigenvalues and eigenvectors are
\begin{align}
\lambda_{\pm}=\pm \sqrt{\epsilon^2+\Delta^2},
\end{align}
\begin{align}
\begin{pmatrix}
u_+ \\ v_+
\end{pmatrix}
&= \mathcal{N}_+
\begin{pmatrix}
\dfrac{1}{\Delta} \left(\epsilon+\sqrt{\epsilon^2+\Delta^2}\,\right)
\\ 1
\end{pmatrix}, \\
\begin{pmatrix}
u_- \\ v_-
\end{pmatrix}
&= \mathcal{N}_-
\begin{pmatrix}
\dfrac{1}{\Delta} \left(\epsilon-\sqrt{\epsilon^2+\Delta^2}\,\right)
\\ 1
\end{pmatrix},
\end{align}
where $\mathcal{N}_\pm$ are the normalization constants.
We can approximate
\begin{align}
\frac{1}{\Delta} \left(\epsilon\pm \sqrt{\epsilon^2+\Delta^2}\,\right)
\sim
\frac{1}{\Delta} (\epsilon \pm |\epsilon|).
\end{align}
Thus, for $\epsilon \geq 0$, we obtain
\begin{align}
\begin{pmatrix}
u_+ \\ v_+
\end{pmatrix}
 \sim
\begin{pmatrix}
1 \\ 0
\end{pmatrix}
,
\quad
\begin{pmatrix}
u_- \\ v_-
\end{pmatrix}
\sim
\begin{pmatrix}
0 \\  1
\end{pmatrix}.
\end{align}
For $\epsilon<0$ on the other hand, because $|\epsilon|=-\epsilon$,
we obtain
\begin{align}
\begin{pmatrix}
u_+ \\ v_+
\end{pmatrix}
 \sim
\begin{pmatrix}
0 \\ 1
\end{pmatrix}
,
\quad
\begin{pmatrix}
u_- \\ v_-
\end{pmatrix}
\sim
\begin{pmatrix}
1 \\  0
\end{pmatrix}.
\end{align}

In summary, in the limit of $|\Delta|/|\epsilon| \ll 1$, for positive (negative) diagonal elements $\epsilon$ of the particle sector in the BdG equations, the negative (positive) eigenvalue mode has $|v|^2 \sim 1$ ($|u|^2 \sim 1$). Regarding the diagonal elements of Eq.~(\ref{Delta}), since $\xi(k, 0) \leq 0$ and $\xi(k\pm K, 0) \geq 0$, the positive (negative) eigenvalue mode for $j=0$ ($j=\pm 1$) band has $|v|^2 \sim 1$.

\end{document}